\documentclass[a4paper]{jpconf}
\usepackage{hyperref}
\usepackage{amsmath}
\usepackage[braket,qm]{qcircuit}
\usepackage{xcolor}

\usepackage{graphicx}
\begin{document}
\title{Studying quantum algorithms for particle track reconstruction in the LUXE experiment}

\author{Lena Funcke$^1$, Tobias Hartung$^{2,3}$, Beate Heinemann$^{4,5}$, Karl Jansen$^6$, Annabel Kropf$^4$, Stefan K\"uhn$^2$, Federico Meloni$^4$, David Spataro$^4$, Cenk T\"uys\"uz$^{6,7}$  and Yee Chinn Yap$^4$}

\address{$^1$ Center for Theoretical Physics, Co-Design Center for Quantum Advantage, and NSF AI Institute for Artificial Intelligence and Fundamental Interactions, Massachusetts Institute of Technology, 77 Massachusetts Avenue, Cambridge, MA 02139, USA}
\address{$^2$ Computation-Based Science and Technology Research Center, The Cyprus Institute, 20 Kavafi Street, 2121 Nicosia, Cyprus}
\address{$^3$ University of Bath, Claverton Down, Bath BA2 7AY, UK}
\address{$^4$ Deutsches Elektronen-Synchrotron DESY, Notkestr. 85, 22607 Hamburg, Germany}
\address{$^5$ Physikalisches Institut, Albert-Ludwigs-Universit\"at Freiburg, Hermann-Herder-Str. 3a, 79104 Freiburg, Germany}
\address{$^6$ Deutsches Elektronen-Synchrotron DESY, Platanenallee 6, 15738 Zeuthen, Germany}
\address{$^7$ Instit\"ut für Physik, Humboldt-Universit\"at zu Berlin, Newtonstr. 15, 12489 Berlin, Germany}

\ead{yee.chinn.yap@desy.de\\ \textnormal{Preprint number:}
  MIT-CTP/5399, DESY-22-027}

\begin{abstract}
  The LUXE experiment (LASER Und XFEL Experiment) is a new experiment in planning at DESY Hamburg, which will study Quantum Electrodynamics (QED) at the strong-field frontier. In this regime, QED is non-perturbative. This manifests itself in the creation of physical electron-positron pairs from the QED vacuum. LUXE intends to measure the positron production rate in this unprecedented regime by using, among others, a silicon tracking detector. The large number of expected positrons traversing the sensitive detector layers results in an extremely challenging combinatorial problem, which can become computationally very hard for classical computers. This paper presents a preliminary study to explore the potential of quantum computers to solve this problem and to reconstruct the positron trajectories from the detector energy deposits. The reconstruction problem is formulated in terms of a quadratic unconstrained binary optimisation. Finally, the results from the quantum simulations are discussed and compared with traditional classical track reconstruction algorithms.
\end{abstract}

\section{Introduction}

LUXE~\cite{CDR} is a proposed experiment at DESY with the aim to study QED in the strong-field regime where QED becomes non-perturbative. The experiment uses the high-energy electron beam from the European XFEL and a high-power laser. Both the interactions of the electron beam with the laser and the interactions of a beam of bremsstrahlung photons with the laser are studied.
The two processes of interest are the Compton scattering process of a photon radiated from the electron in the laser field,
\begin{equation}
  e^- + n \gamma_L \rightarrow e^- + \gamma,
\end{equation}
where $n$ is the number of laser photons $\gamma_L$ participating in the process, and the Breit-Wheeler pair creation
\begin{equation}
  \gamma + n \gamma_L \rightarrow e^+ + e^-
\end{equation}
from the interaction of a photon (which can be the photon from the Compton process) in the laser field.

An important parameter that characterises these interactions is $\xi$, the laser field intensity parameter, defined as
\begin{equation}
  \xi=\sqrt{4\pi\alpha}\,\,\frac{\epsilon_L}{\omega_L m_e}=\frac{m_e \epsilon_L}{\omega_L \epsilon_{cr}},
  \label{eq:xi}
\end{equation}
where $\alpha$ is the fine structure constant, $\epsilon_L$ is the laser field strength, $\omega_L$ is the frequency of the laser, $m_e$ is the electron mass, and $\epsilon_{cr}$ is the critical field strength, also known as the Schwinger limit.

\section{Experimental setup}

\begin{figure}[h]
  \begin{minipage}{18pc}
    \includegraphics[width=18pc]{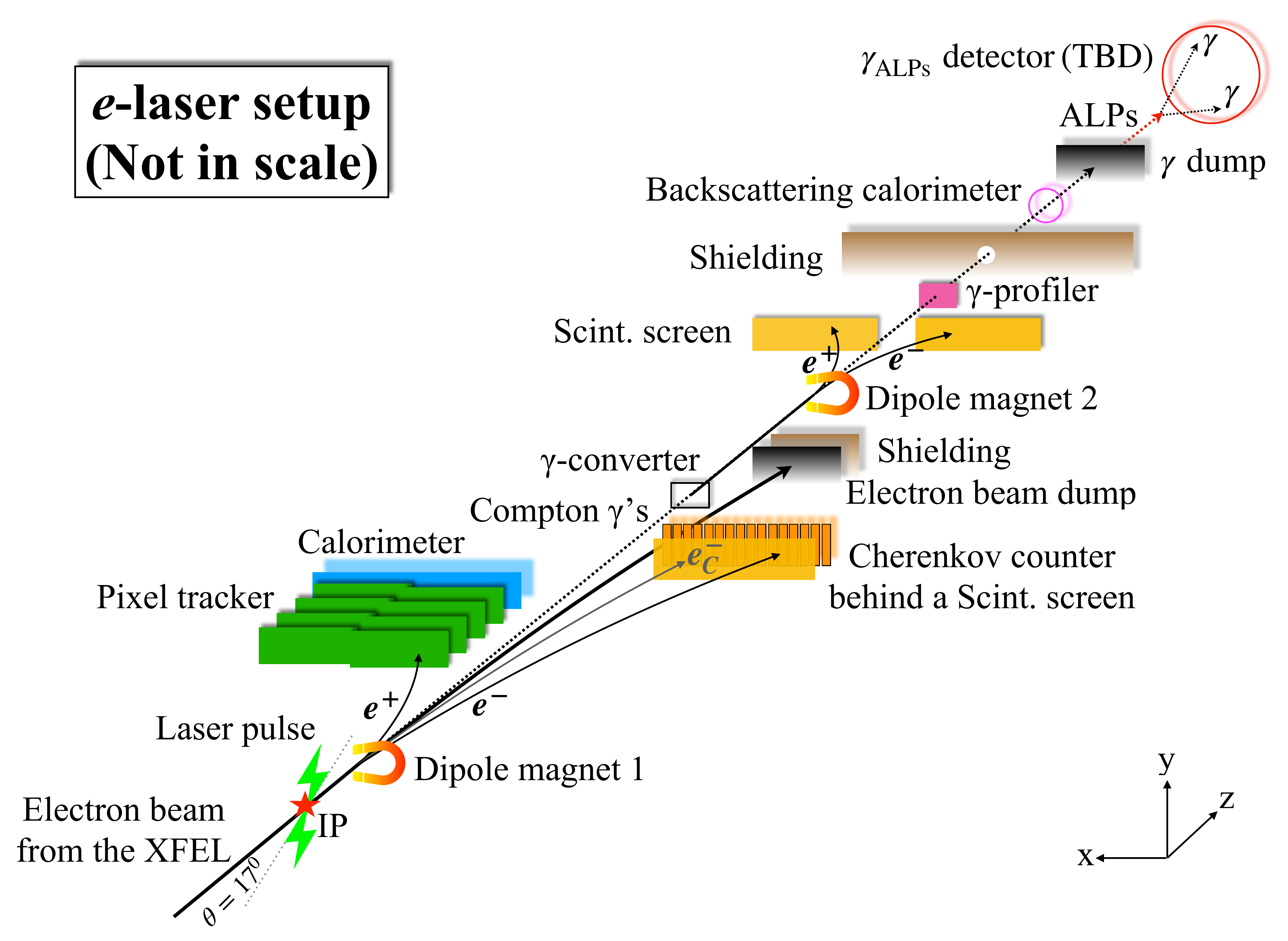}
    \caption{ Schematic layout of LUXE for the e-laser setup. Reproduced from Ref.~\cite{CDR}.}
    \label{fig:LUXEsetup}
  \end{minipage}\hspace{2pc}%
  \begin{minipage}{18pc}
    \includegraphics[width=18pc]{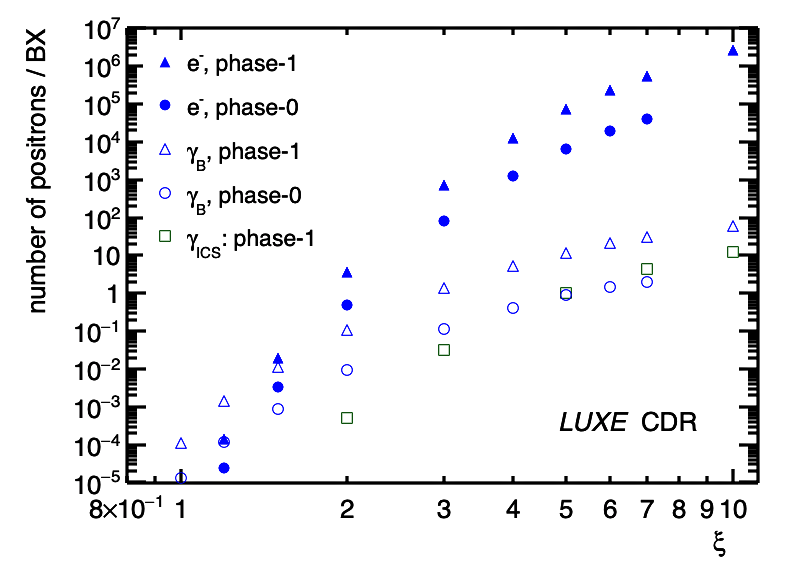}
    %\caption{\label{fig:positronrate}Number of positrons per bunch crossing produced in the e-laser and $\gamma$-laser set-ups for phase-0 and phase-1, as a function of $\xi$ ($\xi_{ \rm nom}$ in the plot). Reproduced from Ref.~\cite{CDR}.}
    \caption{Number of positrons per bunch crossing produced in the e-laser and $\gamma$-laser set-ups for phase-0 and phase-1, as a function of the laser field intensity parameter $\xi$ defined in Eq.~\eqref{eq:xi}. Reproduced from Ref.~\cite{CDR}.}
    \label{fig:positronrate}
  \end{minipage}
\end{figure}

The experimental setup of LUXE in the e-laser mode is shown in Figure \ref{fig:LUXEsetup}. In this setup, the electron beam is guided to the interaction point (IP), where it collides with the laser beam. The initial phase-0 of the experiment will use a 40 TW laser, whereas phase-1 will utilise an upgraded laser power of 350 TW.
The electrons and positrons produced in these interactions are deflected by a magnet and then detected in a variety of detectors. The tracking study presented here concerns mainly positrons, which are detected using a silicon pixel tracking detector. The tracker consists of 4 layers, each comprising two $\approx 27$~cm long staves placed next to each other, which overlap partially, as illustrated in the figure. Each stave contains nine sensors, which each is made up of $512\times1024$ pixels of size $27\times29~{\mu \textrm{m}}^2$.

One of the main measurements at LUXE is the positron flux as a function of the laser field intensity parameter $\xi$ over a large range of $\xi$ values. The positron flux is especially relevant for the e-laser case to measure the Breit-Wheeler pair creation rate without the huge electron beam background. The number of positrons per bunch crossing as a function of $\xi$ spans over ten orders of magnitude, as shown in Figure \ref{fig:positronrate}.
The two main tracking challenges maintain good linearity up to very high multiplicity and keep a very low background rate below $10^{-3}$ per bunch crossing at low $\xi$. To cope with these challenges, we investigate the potential use of quantum computing in track reconstruction. A review of various quantum computing algorithms studied for charged particle tracking can be found in Ref. \cite{GrayReview}.

\section{Data sets and selection}
Simulated data sets are used in this study. The positrons resulting from the signal interactions at the IP, generated using a custom Monte Carlo code named PTARMIGAN~\cite{PTARMIGAN}, are propagated through the dipole magnet and tracking detector using a simplified simulation. In this simulation, four detection layers without gap or overlap are considered and the complexity (position resolution, multiple scattering, etc.) of the simulation can be tuned.

The data set used here corresponds to the e-laser phase-1 setup with $\xi$ values ranging from 3 to 7, and with positron multiplicities between 800 and 500,000. In this study, the tracking problem is limited to the 500 tracks closest to the beamline, such that the size of the problem remains constant but the complexity, due to increasing track density, increases with $\xi$.

The starting point for the tracking is either doublets or triplets, defined as a set of two or three hits in consecutive detector layers. A pre-selection is applied on the initial doublet or triplet candidates to reduce the combinatorial candidates while keeping the efficiency at around 100\%. Doublets are formed first, after applying a pre-selection based on the expected angles from the knowledge of the geometry. Triplets are subsequently constructed by combining doublet candidates with the requirement on the maximum angle difference of the doublet pairs allowed by multiple scattering in the detector. Since triplets consist of three hits, they are formed from either the first to the third layer or from the second to the fourth layer.

\section{Methodology}
\subsection{Classical benchmark}

A tracking based on A Common Tracking Software (ACTS) toolkit~\cite{ACTS} with the combinatorial Kalman Filter (CKF) technique for track finding and fitting is used as a benchmark. In this classical tracking method, track finding starts from seeds, which are the triplets formed from the first three detector layers. An initial estimate of track parameters is obtained from the seed and is used to predict the next hit and is updated progressively, with the measurement search performed at the same time as the fit. Finally, after the track finding and fitting procedures, an ambiguity-solving step is applied to remove tracks with shared hits from the initial track collection.

\subsection{Graph neural network}

Another tracking method explored in this study is based on a graph neural network (GNN)~\cite{HEP.TrkX, Exa.TrkX}. The graph is constructed from doublets, where the hits are nodes and the connections between hits are edges. All nodes of consecutive layers are connected and only the ones that satisfy the pre-selection criteria are kept. Alternating EdgeNetwork and NodeNetwork are applied in the model, such that the model adaptively learns with each iteration which hit connections are important. A hybrid quantum-classical version of the GNN-based tracking also exists~\cite{Q.TrkX}, but is not explored in this work.

\subsection{Quantum approach}

In the quantum approach to tracking, the correct pairs of triplet candidates (where one triplet has hits from the first three layers and the other triplet has hits from the last three layers), which can be combined to form tracks, are identified using a quadratic unconstrained binary optimisation (QUBO), similar to Ref.~\cite{Gray}. The QUBO is expressed as the objective function
\begin{equation}
 O =  \sum_i^N \sum_{j<i} b_{ij} T_i T_j +\sum_{i=1}^{N} a_i T_i ,
 \label{eq:QUBO}
\end{equation}
where $T_i$ and $T_j$ are triplets, $T_i, T_j \in \{0,1\}$, and $a_i$ and $b_{ij}$ are coefficients.

Minimising the QUBO is equivalent to finding the ground state of a Hamiltonian, as explained below. The linear term of the QUBO weighs the individual triplets by their quality quantified by the coefficient $a_i$. The quadratic term expresses the interactions between triplet pairs, where the coefficient $b_{ij}$ characterises the compatibility. The coefficient $b_{ij}$ is positive if the triplets are in conflict, negative if they are compatible to form a track, and zero otherwise.

\begin{figure}[h]
  \begin{center}
    \includegraphics[width=0.7\textwidth]{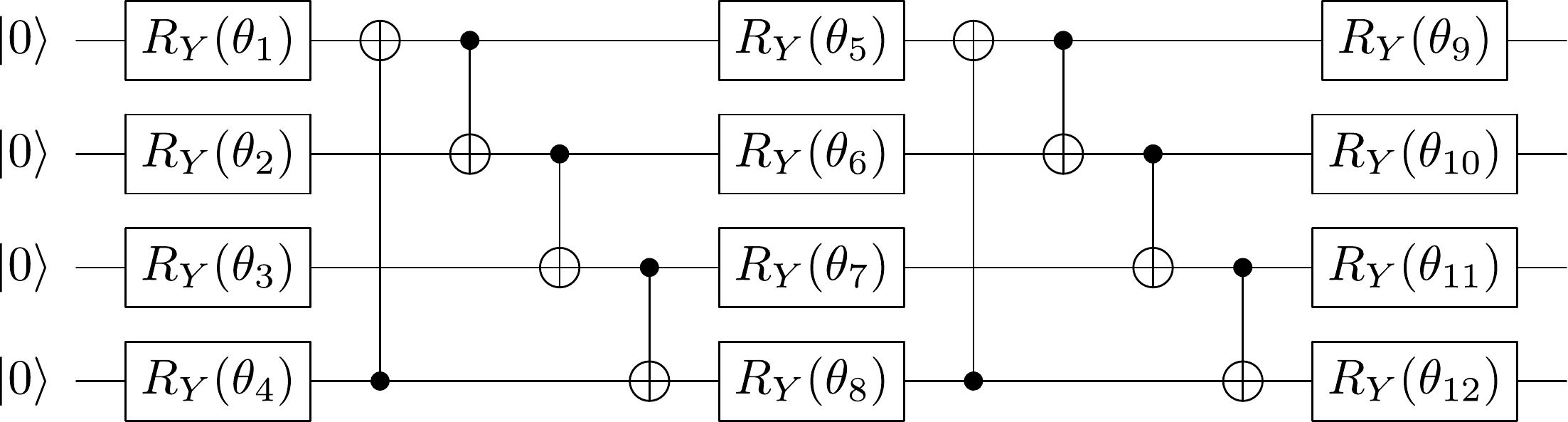}
    %%%%%%%%%%%%%%%%%%%%%%%%%%%%%%%%%%%%%%%
    % Three qubits as pdf version
    %%%%%%%%%%%%%%%%%%%%%%%%%%%%%%%%%%%%%%%
    %\includegraphics[width=0.7\textwidth]{TwoLocalQ3}
    \label{fig:circuit}
    \caption{Layout of the variational quantum circuit using the \textit{TwoLocal} ansatz with $R_Y$ gates and a circular CNOT entangling pattern. For simplicity, only four qubits are shown.}
  \end{center}
\end{figure}

The QUBO in Eq.~\eqref{eq:QUBO} can be mapped to an Ising Hamiltonian and solved using the Variational Quantum Eigensolver (VQE) in Qiskit~\cite{Qiskit}. VQE is a hybrid quantum-classical algorithm to find the minimum eigenvalue of a Hamiltonian. The Ising Hamiltonian
\begin{equation}
  %\mathcal{H}=-\sum_{n=1}^N \sigma_n^x \sigma_{n+1}^x - a \sum_{n=1}^N \sigma_n^x
  \mathcal{H}= -\sum_{n=1}^N\sum_{m<n} \bar{b}_{nm}\sigma_n^x \sigma_{m}^x-\sum_{n=1}^N\bar{a}_n \sigma_n^x 
\end{equation}
has a similar form to the QUBO. An exact solution using the Numpy Eigensolver is available and used as a benchmark. For the VQE, noise is disabled in this study and a simple entangled {\it TwoLocal} ansatz with $R_Y$ gates and a circular CNOT entangler is chosen, as shown in Figure~\ref{fig:circuit}. The selected optimiser is Constrained Optimization by Linear Approximation (COBYLA).

%Figure \ref{} shows the number of interactions a triplet has with other triplets which increases with $\xi$ due to the increasing track density. The numbers of triplet candidates are also given in the figure which increases with $\xi$ despite the constant number of tracks.

%\subsubsection{QUBO solving}

To solve the QUBO, the number of required qubits is determined by the number of triplet candidates. Due to the limited number of qubits available, the QUBO in this work is split into sub-QUBOs of size 7 to be solved iteratively. %The solution of the sub-QUBOs are combined and a tabu search is performed at each iteration. 

\begin{figure}[h]
  \includegraphics[width=38pc]{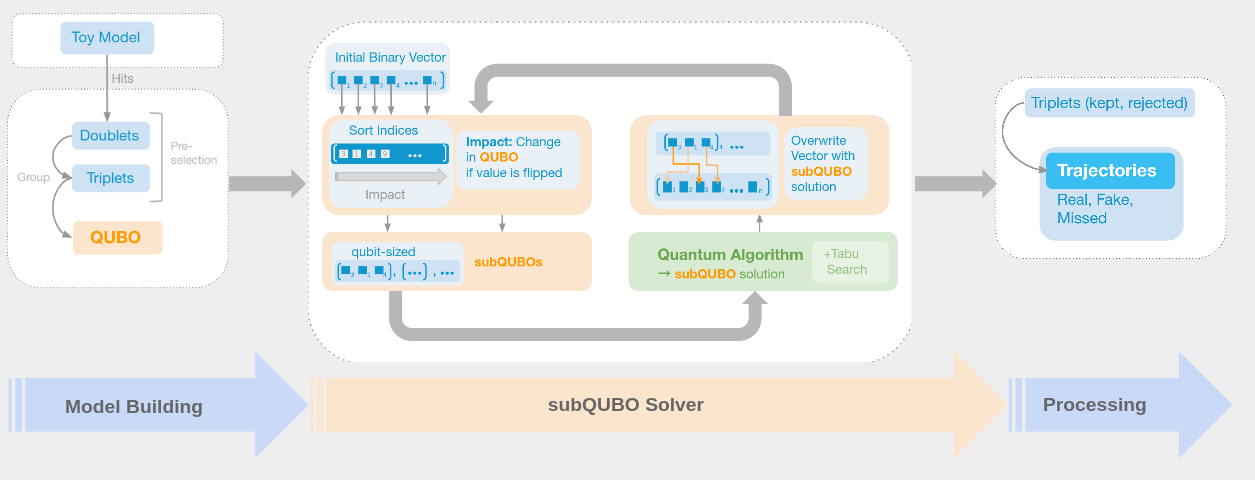}
  \caption{Sketch of the full QUBO solving procedure.}\label{fig:sketch} 
\end{figure}

Figure \ref{fig:sketch} shows a sketch of the QUBO solving process. An initial binary vector is defined by randomly assigning the values \{0,1\} to the triplet candidates. The vector is sorted in order of impact, which is assessed by the change in the value of the QUBO when a bit flip is performed. The splitting into sub-QUBOs is done by partitioning the sorted vector into sub-QUBO size. After the sub-QUBOs are solved, the solution is combined and a tabu search is performed. These steps are repeated for a number of iterations.

%The energy as a function of the number of iterations is shown here on the left for VQE and Numpy Eigensolver for the two cases of scattering on/off. In the scattering case, the Numpy eigensolver is not able to reach the true minimum as defined by the dashed line. We also see a gap between the minimum reached by VQE compared to the one with exact solution.

\section{Results}

The performance of various tracking methods is assessed using the efficiency and the fake rate as metrics, which are computed on the final set of tracks. A track is required to have four hits, which is either found directly with a classical CKF tracking method or by combining selected triplet pairs into quadruplets. A track is only considered matched if the track has all four hits matched to the same particle.

The efficiency and fake rate are defined as
\begin{equation}
  \textrm{Efficiency} = \frac{N_{\rm tracks}^{\rm matched}}{N_{\rm tracks}^{\rm generated}} \qquad \text{and}\qquad
  \textrm{Fake rate} = \frac{N_{\rm tracks}^{\rm fake}}{N_{\rm tracks}^{\rm reconstructed}}\, .
\end{equation}

\begin{figure}[h]
  \begin{minipage}{18pc}
    \includegraphics[width=18pc]{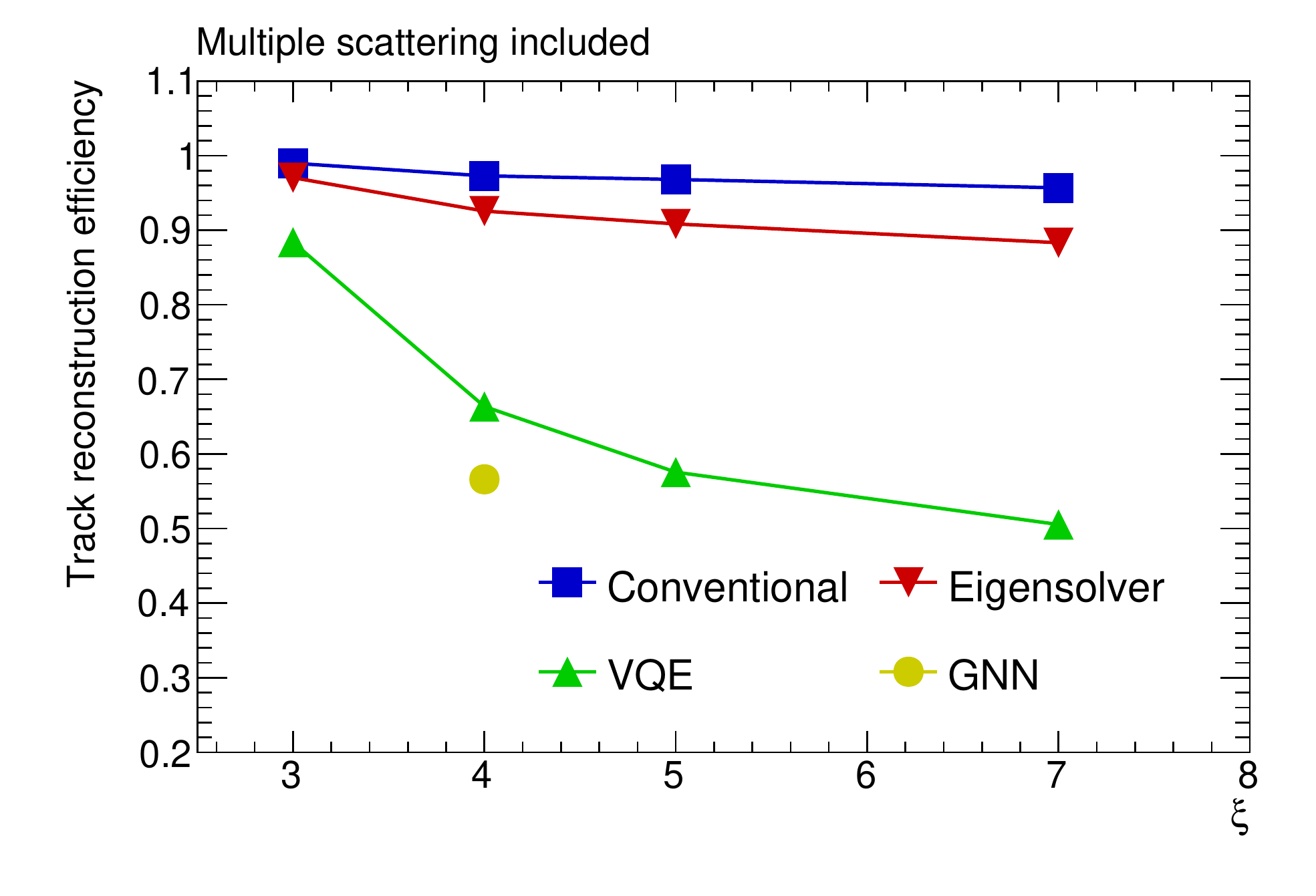}
    \caption{Track reconstruction efficiency as a function of the field intensity parameter $\xi$.}\label{fig:eff} 
  \end{minipage}\hspace{2pc}%
  \begin{minipage}{18pc}
    \includegraphics[width=18pc]{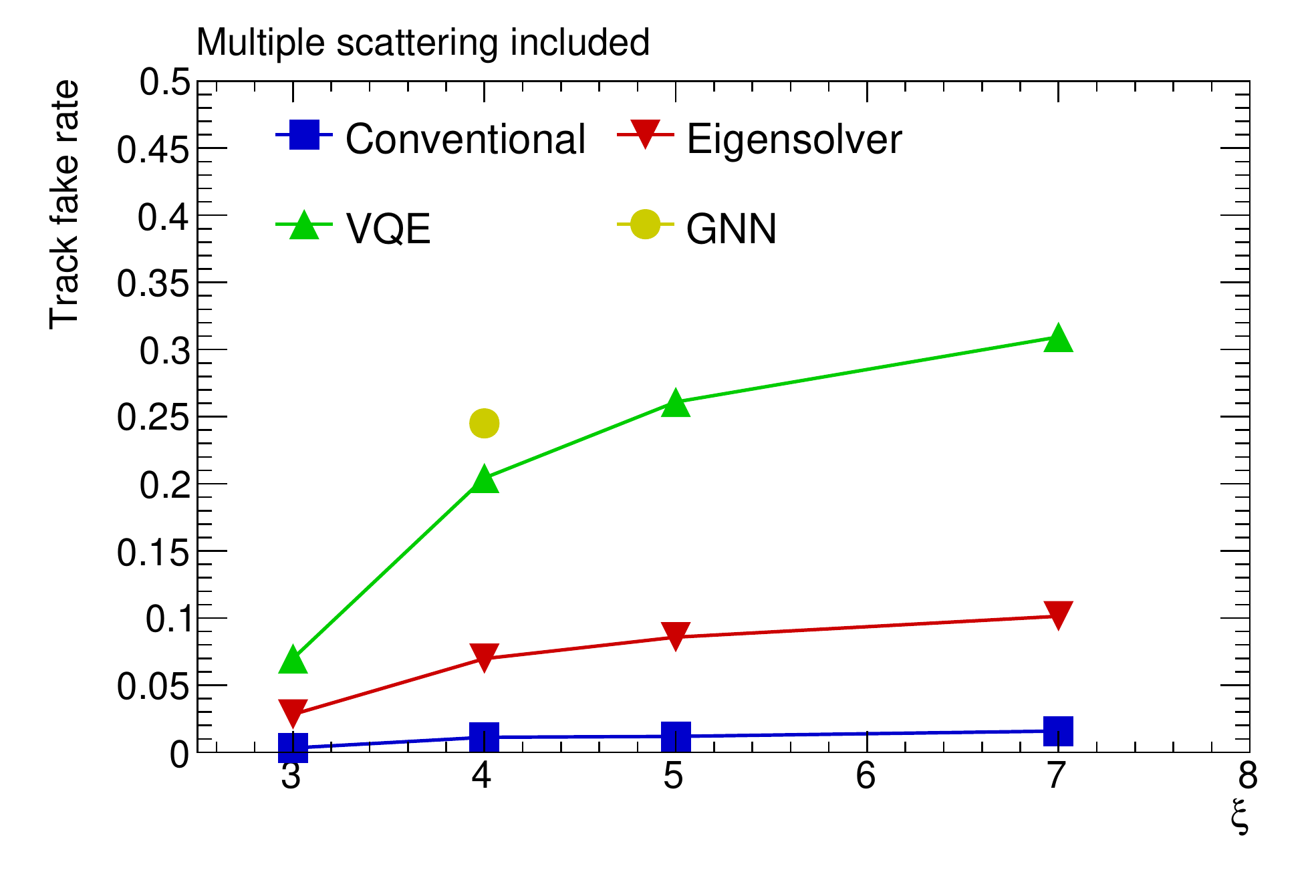}
    \caption{Track fake rate as a function of $\xi$.}\label{fig:fakerate}
  \end{minipage}
\end{figure}

Figure \ref{fig:eff} and \ref{fig:fakerate} show the track reconstruction efficiency and fake rate as a function of the laser field intensity parameter $\xi$ for the four methods tested: conventional CKF-based tracking, GNN-based tracking, VQE, and the VQE exact solution using the Eigensolver.

The conventional CKF-based tracking, while performant, deteriorates with $\xi$. The performance of CKF-based tracking is used as a benchmark to demonstrate the performance that can be realistically achieved. The intial results using the Eigensolver are slightly poorer than the CKF tracking, which thus need to be further optimised. The results for VQE demonstrate that our initial implementation is less effective; however, it can also be further optimised, e.g., by using a more appropriate choice of circuit ansatz and optimiser. The limited size of the quantum device, which prompts the sub-QUBO algorithm, is also a potential contributing factor to the initial degradation of the quantum approach. We will study these different factors in detail in future work. Finally, the preliminary result for the GNN-based tracking is shown for a specific $\xi$ value of 4. Here, the current underperformance is likely caused by a lack of statistics in the training sample, which will also be extended and further optimised in future work. 

\section{Conclusion}

The use of a hybrid quantum-classical algorithm in track reconstruction is studied along with a conventional tracking method as well as a GNN-based tracking. A first implementation of track reconstruction in the LUXE experiment using quantum devices is in place. 
%Preliminary results suggest that the performance is encouraging but limited by the size of the quantum device. The results for VQE show that our initial implementation 
%
In its current version, the performance is less effective than the conventional tracking method, which implies that the quantum algorithm needs to be further optimised, in particular by improving the circuit ansatz. Moreover, the performance of the sub-QUBO algorithm is currently limited by the size of the quantum device. We plan to mitigate these effects in extended future studies.

\ack{The work by B.H., A.K., F.M., D.S. and Y.Y. was in part funded by the Helmholtz Association - “Innopool Project LUXE-QED”. K.J. and C.T. are supported in part by the Helmholtz Association - “Innopool Project Variational Quantum Computer Simulations (VQCS)”. L.F.\ is supported by the U.S.\ Department of Energy, Office of Science, National Quantum Information Science Research Centers, Co-design Center for Quantum Advantage (C$^2$QA) under contract number DE-SC0012704, by the DOE QuantiSED Consortium under subcontract number 675352, by the National Science Foundation under Cooperative Agreement PHY-2019786 (The NSF AI Institute for Artificial Intelligence and Fundamental Interactions, http://iaifi.org/), and by the U.S.\ Department of Energy, Office of Science, Office of Nuclear Physics under grant contract numbers DE-SC0011090 and DE-SC0021006. S.K.\ acknowledges financial support from the Cyprus Research and Innovation Foundation under project ``Future-proofing Scientific Applications for the Supercomputers of Tomorrow (FAST)'', contract no.\ COMPLEMENTARY/0916/0048. This work has benefited from computing services provided by the German National Analysis Facility (NAF).}


\section*{References}
\begin{thebibliography}{9}
  \bibitem{CDR} H.\ Abramowicz et al., Conceptual design report for the LUXE experiment, {\it Eur. Phys. J. ST} {\bf 230}, 2445 (2021), {\it Preprint} 2102.02032
  \bibitem{GrayReview} H.\ Gray, Quantum pattern recognition algorithms for charged particle tracking, {\it Phil. Trans. R. Soc. A.} {\bf 380}, 20210103 (2021)
  \bibitem{PTARMIGAN} T.\ G.\ Blackburn, A.\ J.\ MacLeod, and B.\ King, From local to nonlocal: higher fidelity simulations of photon emission in intense laser pulses, {\it New J. Phys.} {\bf 23}, 085008 (2021), {\it Preprint} 2103.06673
  \bibitem{ACTS} X.\ Ai et al., A Common Tracking Software Project, {\it Preprint} 2106.13593 (2021)
  \bibitem{HEP.TrkX} S.\ Farrell et al., Novel deep learning methods for track reconstruction, {\it Preprint} 1810.06111 (2018)
  \bibitem{Exa.TrkX} X.\ Ju et al., Performance of a geometric deep learning pipeline for HL-LHC particle tracking {\it Eur. Phys. J. C} {\bf 81}, 876 (2021), {\it Preprint} 2103.06995
  \bibitem{Q.TrkX} C.\ T\"uys\"uz, C.\ Rieger, K.\ Novotny, B.\ Demirk\"oz, D.\ Dobos, K.\ Potamianos, S.\ Vallecorsa, J.\ Vlimant, and R.\ Forster, Hybrid Quantum Classical Graph Neural Networks for Particle Track Reconstruction, {\it Quantum Mach.\ Intell.} {\bf 3}, 29 (2021), {\it Preprint} 2109.12636
  \bibitem{Gray} F.\ Bapst, W.\ Bhimji, P.\ Calafiura, H.\ Gray, W.\ Lavrijsen, L.\ Lindner, and A.\ Smith, A Pattern Recognition Algorithm for Quantum Annealers, {\it Comput.\ Softw.\ Big Sci.} {\bf 4} 1 (2020)
  \bibitem{Qiskit} M.\ Treinish et al., Qiskit: An Open-source Framework for Quantum Computing, \href{https://doi.org/10.5281/zenodo.2573505}{\it Zenodo} (2022)
\end{thebibliography}
\end{document}